\documentclass[%
 aip,
 amsmath,amssymb,
 reprint,%
]{revtex4-2}

\usepackage{graphicx}
\usepackage{dcolumn}
\usepackage{bm}

\usepackage[utf8]{inputenc}
\usepackage[T1]{fontenc}
\usepackage{mathptmx}
\usepackage{physics}
\usepackage[disable]{todonotes}
\usepackage{parskip}
\usepackage{mhchem}
\usepackage{siunitx}
\DeclareSIUnit\gauss{G}
\DeclareSIUnit\torr{Torr}
\DeclareSIUnit\bar{bar}
\usepackage{float}
\DeclareSIUnit\gauss{G}

\begin{document}

\preprint{AIP/123-QED}

\title[]{Rapid Electromagnetic Induction Imaging with an Optically Raster-Scanned Atomic Magnetometer}

\author{B. Maddox}
\affiliation{%
Department of Physics and Astronomy, University College London, Gower Street, London WC1E 6BT,
United Kingdom
}%
\author{C. Deans}%

\affiliation{%
Department of Physics and Astronomy, University College London, Gower Street, London WC1E 6BT,
United Kingdom
}%
\affiliation{%
Present address: UKRI National Quantum Computing Centre, Harwell Campus, Didcot OX11 0GD, United Kingdom
}%

\author{H. Yao}
\author{Y. Cohen}
\author{F. Renzoni} 
\email{f.renzoni@ucl.ac.uk}
\affiliation{%
Department of Physics and Astronomy, University College London, Gower Street, London WC1E 6BT,
United Kingdom
}%

\date{\today}

\begin{abstract}
We present an apparatus to overcome the limitations of mechanical raster-scanning in electromagnetic induction imaging (EMI) techniques by instead performing a 2D optical raster-scan within the vapour cell of a radio-frequency atomic magnetometer (RF-AM). A large cuboidal \ce{^{87}Rb} vapour cell is employed to act as the medium of an RF-AM with the pump and probe beams translated in the cell via acousto-optics. The technique is shown to give robust and repeatable magnetic measurements over the cell volume and successfully resolves conductive targets with EMI. Optical raster-scanning removes the limitation of slow mechanical actuation and a fast imaging procedure is enacted resolving conductive targets at a rate of \SI{40}{\milli \second / pixel}.       
\end{abstract}

\maketitle
Electromagnetic induction imaging\cite{Griffiths_2001} (EMI) invites the potential for a non-contact non-destructive inherently safe imaging technique suitable for applications in biomedicine, security and surveillance, and industrial monitoring\cite{deans2016electromagnetic,deans2020sub,bevington2018non}. The use of an oscillating magnetic field (primary field) to induce eddy currents in the target material generates a reciprocal oscillating magnetic field (secondary field) with properties that reveal the electrical and magnetic characteristics of the material. The secondary field can then be measured with a magnetic sensor and atomic magnetometers (AMs) offer the potential to unlock extreme sensitivity in the low-frequency regime\cite{budker2007optical,ledbetter2007detection,savukov2007detection}, with immediate applications in  through-barrier  imaging\cite{deans2017through,portable2}
and potential applications in human organ screening\cite{marmugi_SR}. 
Using EMI to image conductive objects with AMs has been successful in several previous experiments \cite{deans2020sub,wickenbrock2016eddy,bevington2018non,bevington2019enhanced,jensen2019detection, rushton2022unshielded} but so far all previous implementations have required either moving the object or the sensor. Moving the object can be impossible in many applications. On the other hand, moving the sensor, while experimentally demonstrated \cite{portable1,portable2,portable3}, poses limitations in the stability and repeatability of the AM measurement. In both cases, the speed at which an image can be taken is limited by the speed of movement from pixel to pixel which can be on the order of seconds when using mechanical actuators.
\begin{figure}
    \centering
    \includegraphics[width=\linewidth]{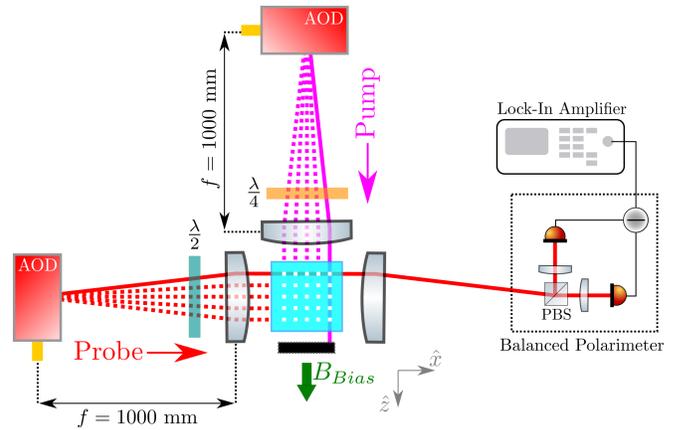}
    \caption{Schematic of the optical setup for the raster-scanned atomic magnetometer. The pixel array is simplified to a $5 \times 5$ matrix with white squares to represent the pixel position in the cell. Solid red lines show the paths of the pump and probe beams in a single pixel measurement whereas the dashed lines represent the beam paths used to address the subsequent pixels.  }
    \label{FIG:Optical_Schematic}
\end{figure}
\begin{figure}[h]
    \centering
   \includegraphics[width=\linewidth]{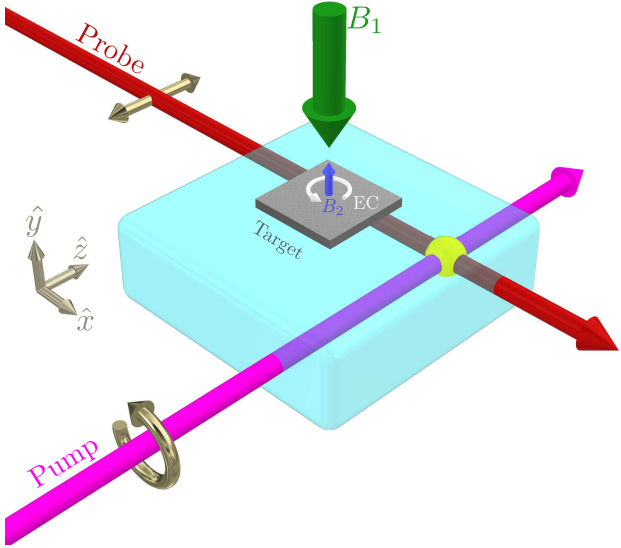}
    \caption{Illustration of EMI utilising the optically raster-scanned atomic magnetometer. The primary field ($B_1$) is applied to the target, inducing eddy currents (EC) which generate a secondary field ($B_2$). The sum of the magnetic fields can then be measured locally at the pump-probe intersection (yellow sphere) where the pump (magenta) and probe (red) overlap in the cell. Golden arrows along the beam axes signify the light polarisation.}
    \label{FIG:Optical_Schematic_side}
\end{figure}
\begin{figure}[h!]
    \centering
    \includegraphics[width=.95\linewidth]{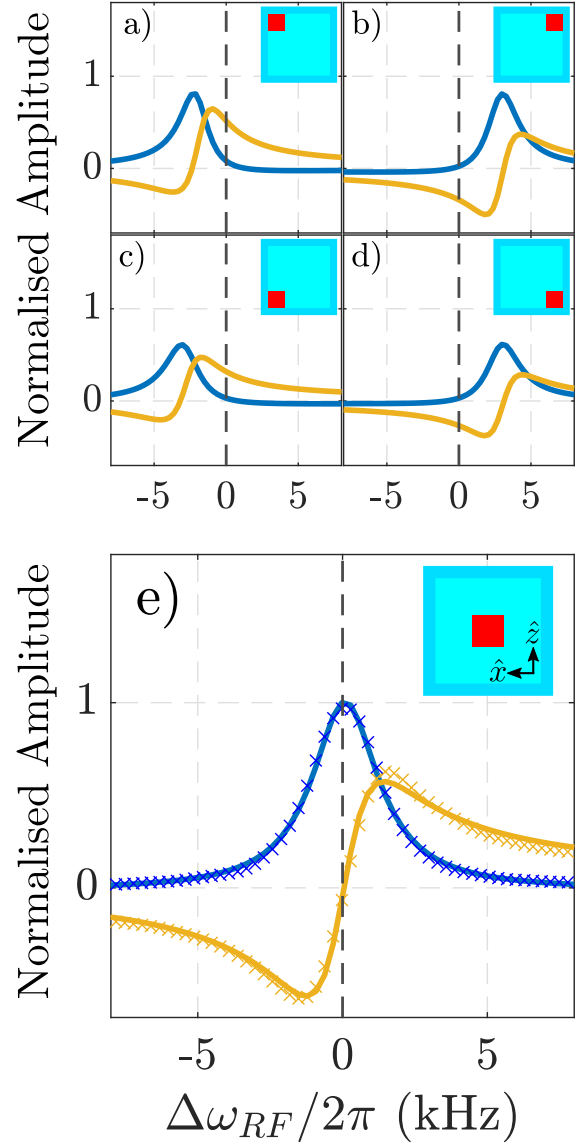}
    \caption{Magnetic resonance lineshapes as $\omega_{RF}$ is swept through the resonance for the four corners (a-d) and centre e) of the $40{\times}\SI{40}{\mm^2}$ imaging area. The blue and yellow crosses show the data from the X and Y outputs of the LIA respectively and solid lines represent the Lorentzian and dispersive lineshapes that were fitted to the LIA data. The amplitudes of all plots are normalised to the maximum of the lorentzian fit in the the central resonance in e). The inset shows a graphic to signify the relative position in the cell with the experimental $x$-$z$ plane indicated in the inset of e). Data is not shown in a-d to ease examination.}
    \label{FIG:Resonance_Grid}
\end{figure}

Here we present an apparatus to overcome these issues by instead optically moving the sensor volume within the atomic vapour cell. Deflection of the pump and probe beams via Acousto-Optical Deflectors (AODs) translates the beams within the vapour cell and allows the movement of the pump-probe intersection. This method circumvents the need for any moving parts, with the speed of sensor movement limited only by the rise time of the AOD (${<}\SI{10}{\micro \second}$). The minimum pixel size is limited only by the diffusion length of the $\ce{^{87}Rb}$ atoms through the sensor volume (\SI{1.95}{\mm} for our experimental conditions), with no restriction on spacing between pixels. The method allows for a sequential measurement of a tightly spaced AM array, without the complexity of numerous polarimeter setups and high optical power that would be required for a concurrent array measurement.  

Fig \ref{FIG:Optical_Schematic} shows a schematic of the optical setup of the magnetometer. 
The magnetometer interaction scheme is similar to previous implementations of radio-frequency (RF) magnetometers \cite{deans2018sub,chalupczak2012room}. A room-temperature cuboidal isotopically-enriched $\ce{^{87}Rb}$ vapour cell of dimensions $60 {\times} 60 {\times} \SI{20}{\mm^3}$ (W$\times$L$\times$H) acts as the atomic medium with \SI{500}{\torr} of $\ce{He}$ buffer gas and \SI{100}{\torr} of $\ce{N_2}$ quenching gas. The pump laser is detuned +\SI{1.8}{\giga \hertz} from the $\ket{F=2} \rightarrow  \ket{F'} $ transition on the $D_1$ line of $\ce{^{87}Rb}$. The probe laser is detuned by +\SI{7.6}{\giga \hertz} from the $\ket{F=1} \rightarrow  \ket{F'} $ transition on the $D_2$ line of $\ce{^{87}Rb}$. The probe is linearly polarised, has a $1/e^2$ diameter of \SI{2.5}{\mm} in the $x$-$z$ plane and propagates along $\hat{x}$, while the pump is circularly polarised, has a $1/e^2$ diameter of \SI{3.4}{\mm} in the $x$-$z$ plane and propagates along $\hat{z}$. A cubic three-axis Helmholtz coil system, centred on the vapour cell, creates the magnetic fields necessary for the RF magnetometer. A DC magnetic field in the $\hat{z}$ direction induces a bias field of $\mathbf{B}_{Bias}= \SI{150}{\milli \gauss}$ which creates Zeeman splitting of the $F=2$ hyperfine manifold, setting the magnetometer resonance at $\omega_0 = 2\pi\times$\SI{105}{\kilo \hertz}. The system is unshielded with the aforementioned three pairs of square Helmholtz coils, orthogonally aligned and centred on the cell, providing magnetic compensation. An array of 4 fluxgate sensors (Stefan-Mayer FLC100) arranged in a rectangular formation centred on the vapor cell, sits in the $x$-$z$ plane ($155{\times}\SI{115}{\mm^2}$) and measures the bias field in the $\hat{z}$ direction. The use of an array of fluxgate sensors allows the inference of the magnetic field at the centre of the cell without requiring spatial access \cite{Mag_array_2,Mag_array_1}. The fluxgate array measurement is then stabilised via a proportional-integral-derivative (PID) controller (SRS SIM960) that servos the bias coil current.  The Helmholtz coils in the $\hat{x}$ and $\hat{y}$ directions are used to compensate for ambient magnetic fields and to align  $\mathbf{B}_{Bias}$ to $\hat{z}$. A square RF coil generates an AC magnetic field $\mathbf{B}_{RF}$ that oscillates at a frequency $\omega_{RF}$ in the $\hat{y}$ direction. With dimensions $55{\times}\SI{55}{\mm^2}$, the RF coil is larger than the imaging area to provide a near-uniform field and is positioned \SI{25}{\mm} above the cell allowing space for the target object to be placed. The probe beam interacts with the atomic medium before its polarisation is measured by a polarimeter, formed by a polarising beam-splitter (PBS) and a balanced photodetector. $\mathbf{B}_{RF}$ drives the Larmor precession of the atoms about $\mathbf{B}_{Bias}$ imprinting a polarisation oscillation onto the probe which is then read out by a dual-phase lock-in amplifier referenced at $\omega_{RF}$.  

The buffer gas reduces the mean free path of the $\ce{^{87}Rb}$ atoms such that two vital conditions are achieved. Firstly, the atom-light interaction time is increased leading to a longer transverse relaxation time which increases the sensitivity of the magnetometer. Secondly, since the motion of the atoms is restricted, the detected RF field in a measurement pertains only to atoms that diffuse across the volume of intersection between the pump and probe beams during a measurement cycle. 

The movement of the pump and probe beams is achieved via AODs (AA DTSX-400-780) which diffract the beams by an angle $\theta_B$ according to the Bragg condition 
\begin{equation}
\label{EQ:Bragg_angle}
 \sin{\theta_B}= m \frac{\lambda}{2 n \Lambda} 
\end{equation}
where $\lambda$ is the wavelength of the incoming light, $\Lambda$ is the speed of the sound wave in the AOD, $n$ is the refractive index of the AOD crystal and $m$ is the order of diffraction. Both beams enter their own respective AODs and the first order of diffraction is taken as the emergent beam, with the other orders blocked. The small angle $\alpha$ between the first order and the zeroth order beams is then given by 
\begin{equation}
\label{EQ:Bragg_angle_2}
 \alpha = \frac{\lambda \nu}{ \Lambda} 
\end{equation}
where $\nu$ is the frequency of the signal driving the AOD. The first order beam is then incident on an $f= \SI{1000}{\mm}$ plano-convex lens, which converts the angular deviation of the beam to a translation across the cell. Applying the specific parameters of the system to equation \ref{EQ:Bragg_angle} we find that for small angles, the translation of the beam position $x$ in the cell $dx/d\nu = f \lambda / \Lambda = \SI{1.2}{\mm / \mega \hertz}$. Sweeping an AOD driving frequency range of $\Delta \nu = \SI{50}{\mega \hertz}$ gives a full raster scan of the beam across the cell width. We have experimentally confirmed this by measuring the movement of the beam after the lens. The position of pump-probe intersection in the cell can then be swept in a 2D array in the $x$-$z$ plane, with each intersection corresponding to a pixel of the EMI image. The size of the array can be made to be any $n \times n$ array.
For each pixel measurement, $\omega_{RF}$ is swept through \mbox{ $\omega_{0} - 5 \Gamma \rightarrow \omega_{0} + 5\Gamma$} where $\Gamma= 2 \pi \times$\SI{2.4}{\kilo \hertz} is the linewidth (full width at half maximum) of the magnetic resonance. The dual-phase lock-in amplifier (LIA) extracts the in-phase $X(\omega)$ and quadrature $Y(\omega)$ components oscillating at $\omega_{RF}$. X is fitted to a Lorentzian profile while Y is fitted to a dispersive profile. The radius $R(\omega)=\sqrt{X^2 + Y^2}$ and phase \mbox{$\phi(\omega)=\arctan(X/Y)$} are then calculated with the fitted parameters of X and Y, for each pixel.

\begin{figure}
    \centering
    \includegraphics[width=.9\linewidth]{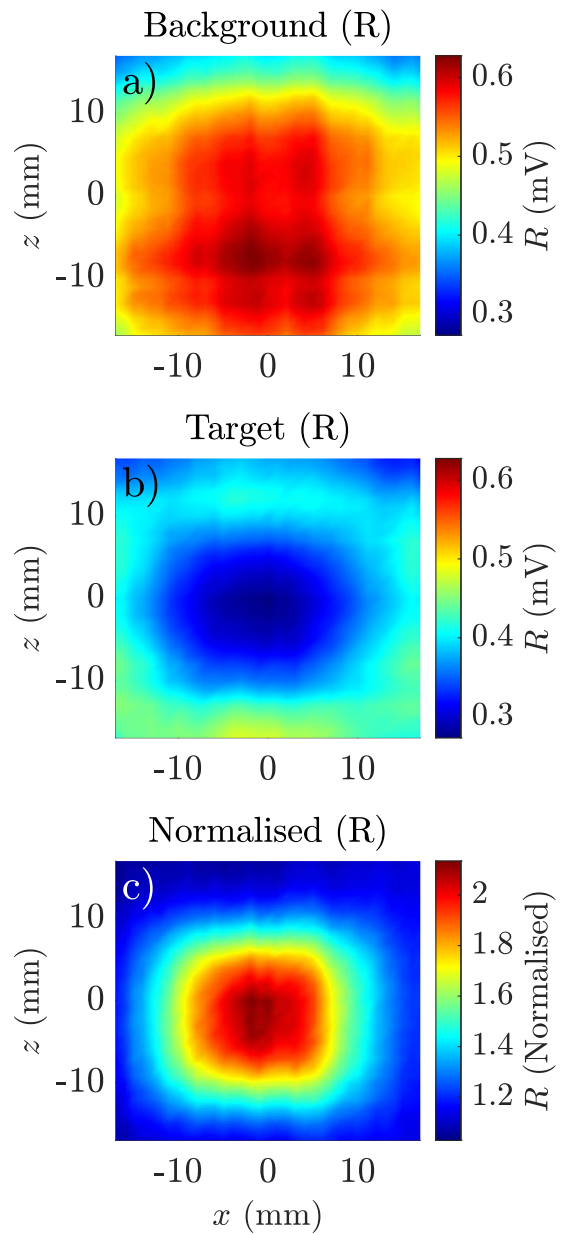}
    \caption{$R$ images of the background (a), the target (b) and the normalised image (c) of an Cu square with dimensions 15$\times$15$\times$\SI{2}{\mm^3}. A pixel array of $35{\times}35$ was used for each image, with \SI{1}{\mm} step per pixel.}
    \label{FIG:COPPER_SQUARE_STRIP}. 
\end{figure}
\begin{figure}[h]
    \centering
    \includegraphics[width=\linewidth]{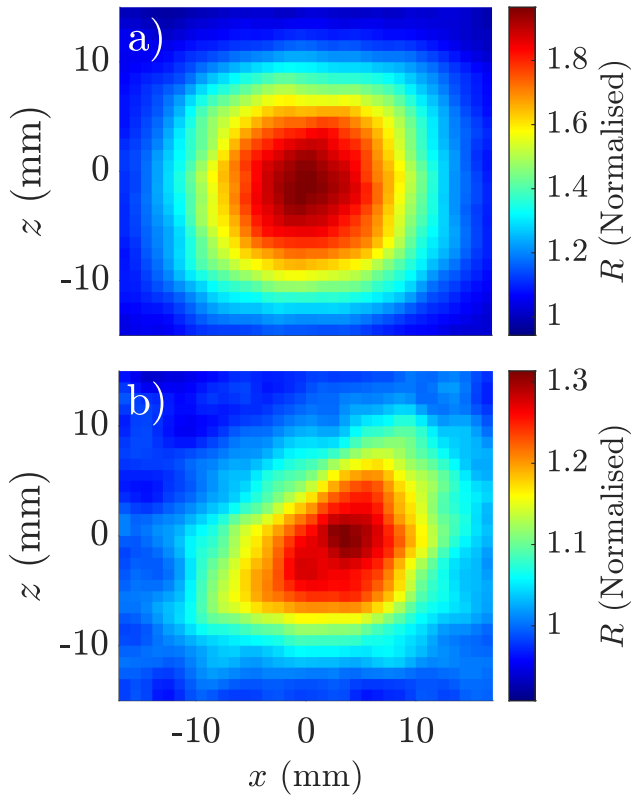}
    \caption{Fast images of a Cu square a) and a Cu triangle b) using the fast EMI procedure with a measurement duration of \SI{40}{\milli \second / pixel}. Both images are normalised with the background as in FIG \ref{FIG:COPPER_SQUARE_STRIP}c and are on the same spatial scale. A pixel array of $35{\times}35$ was used for each image, with \SI{1}{\mm} step per pixel. The images are smoothed by a convolution of the dataset with a nearest-neighbour Gaussian filter of radius 1 pixel.}
    \label{FIG:FAST_IMAGES}
\end{figure}

FIG \ref{FIG:Resonance_Grid} shows the recorded $X$ and $Y$ traces for the centre and corner positions of the whole imaging area, demonstrating the ability to achieve a magnetic resonance across the cell. The resonant frequencies of the outer pixels shift relative to the central pixel resonance showing that $\mathbf{B}_{Bias}$ is inhomogeneous over the imaging area. This is circumvented by tracking the whole resonance for each pixel and extracting the resonant frequency from the Lorentzian lineshape fitting. The frequency distribution across the imaging area of $ \omega_0 \pm2\pi\times\SI{2}{\kilo \hertz}$ gives a negligible change in the skin-depth of the Cu targets $(201\pm1)\SI{}{\micro \meter}$. The amplitude of the magnetic resonances across the imaging area is also inhomogeneous (further reinforced in FIG \ref{FIG:COPPER_SQUARE_STRIP}a) necessitating a prior background measurement to accommodate for this. \newline The system is then set to scan an area of $\SI{38}{\mm} \times \SI{38}{\mm}$ in increments of \SI{1}{\mm}. A $\SI{25}{\mm} \times \SI{25}{\mm}$ copper square with a thickness of \SI{1}{\mm} acts as the target and is placed between the RF coil and the cell as illustrated in FIG \ref{FIG:Optical_Schematic_side}. Images are taken with, and without, the target. Normalising the target image by the background image accommodates for the inhomogeneity of the magnetometer amplitude across the scanning area, where a decreased amplitude would show a decreased efficiency for measuring a change in field. The normalisation is done by dividing each background pixel by the corresponding target pixel, leading to the normalised image. Typically, owing to the repeatability of the system, only one background image is required with no need to repeat this from target to target. FIG 4 shows the target, background and normalised images for the copper square target. While the effect of the  Cu square can be seen in the target image, higher contrast is achieved when normalizing to the background. 

The pixel duration, can be broken down into three distinct periods: the optical translation of the pump and probe beams to the position of the new pixel, the computer control of the instruments to set up for the measurement and the atomic measurement duration itself. The use of the AODs significantly reduces the time spent rastering through the pixels compared to mechanical actuation (${\sim} \SI{1}{ \second / pixel}  \rightarrow {<}\SI{10}{ \micro \second / pixel}$) rending this phase negligibly short. For computer control, a LabVIEW script is implemented to run the experiment and is limited by the computer iteration loop on the order of ${\sim}\SI{100}{\milli \second / pixel}$. However, sequencing the control events in hardware can allow control signals to be pre-programmed and limited to the DAC (digital-to-analogue converter) risetime ${<}\SI{1}{\micro \second / pixel}$, effectively removing this latency. Ultimately this leaves the measurement duration as the main limiting factor on imaging speed. Previously, as in FIG \ref{FIG:COPPER_SQUARE_STRIP}, the full resonance is swept per pixel with a full LIA measurement needed for each value of $\omega_{RF}$ in the sweep. To speed up the imaging process this can be reduced to 1 point, at the maximum of the resonance. The repeatability of the bias field stabilisation allows for a predictable resonant frequency at each pixel, allowing $\omega_{RF}$ to be predetermined from the fitted resonances in the background image. FIG \ref{FIG:FAST_IMAGES} shows the employment of this procedure, to resolve a Cu square ($25{\times}\SI{25}{\mm^2}$) and a Cu right-angled triangle (of side length \SI{25}{\mm}). Utilising this scheme allows a measurement duration of \SI{40}{\milli \second / pixel}. 

The method outlined in this article enables fast EMI measurements in a sequential array without the complexity required for a concurrent array measurement. A concurrent array measurement would require a balanced detector array with the number of pixels matching the resolution of the EMI image. To replicate the image in FIG \ref{FIG:COPPER_SQUARE_STRIP} this requirement would be on the order of thousands of balanced detectors, which can be challenging to produce with very low noise and low crosstalk between pixels. The sequential array method also allows for adjustments at each pixel, where inhomogeneity in the working conditions (for example bias field) can be accommodated for. In principle, the imaging size is only limited by the dimensions of the atomic medium. Glassblown cells or vacuum systems can be engineered to the meter scale, enabling an imaging size that would for example, be appropriate for luggage screening. Programming the experimental sequence to hardware would effectively remove the latency of control electronics allowing the imaging speed to be reduced to the measurement duration. Improvements in the magnetic sensitivity will allow for further reductions in the required measurement duration.    

\begin{acknowledgments}
This work has been partially funded by the Future Aviation Security Solutions (FASS) programme, a joint Department for Transport and Home Office initiative, under Contract No. DSTLX-1000140529. With thanks for the technical oversight and programme management provided by the Defence Science and Technology Laboratory (Dstl). This research has also received funding from the UK Engineering and Physical Sciences Research Council (EPSRC)  (Grant No. EP/R511638/1).
\end{acknowledgments}

\bibliographystyle{aipnum4-1}
\bibliography{aipsamp}

\end{document}